\documentclass[aps,pre,epsf,superscriptaddress,amsmath,amssymb,amsfonts,twocolumn,showpacs]{revtex4-1}

\usepackage{graphicx}
\usepackage{epsfig}
\usepackage{dcolumn}
\usepackage{bm}
\usepackage{braket}
\usepackage{amsmath}
\usepackage{color}

\begin{document}
\title{Quantum dynamical response of ultracold few boson ensembles\\ in finite optical lattices to multiple interaction quenches}

\author{J. Neuhaus-Steinmetz}
\affiliation{Zentrum f\"{u}r Optische Quantentechnologien,
Universit\"{a}t Hamburg, Luruper Chaussee 149, 22761 Hamburg,
Germany}
\author{S.I. Mistakidis}
\affiliation{Zentrum f\"{u}r Optische Quantentechnologien,
Universit\"{a}t Hamburg, Luruper Chaussee 149, 22761 Hamburg,
Germany}
\author{P. Schmelcher}
\affiliation{Zentrum f\"{u}r Optische Quantentechnologien,
Universit\"{a}t Hamburg, Luruper Chaussee 149, 22761 Hamburg,
Germany} \affiliation{The Hamburg Centre for Ultrafast Imaging,
Universit\"{a}t Hamburg, Luruper Chaussee 149, 22761 Hamburg,
Germany}

\date{\today}

\begin{abstract}

The correlated non-equilibrium quantum dynamics following a multiple interaction quench protocol for few-bosonic ensembles confined in 
finite optical lattices is investigated.  
The quenches give rise to an interwell tunneling and excite the cradle and a breathing mode. 
Several tunneling pathways open during the time interval of increased interactions, while only a few occur when the system is quenched back to its original interaction strength.   
The cradle mode, however, persists during and in between the quenches, while the breathing mode possesses dinstinct frequencies. 
The occupation of excited bands is explored in detail revealing a monotonic behavior with increasing quench amplitude 
and a non-linear dependence on the duration of the application of the quenched interaction strength.  
Finally, a periodic population transfer between momenta for quenches of increasing interaction is observed, with a power-law frequency dependence on the quench amplitude. 
Our results open the possibility to dynamically manipulate various excited modes of the bosonic system.   
\end{abstract}

\maketitle

\section{Introduction}

Ultracold atoms in optical lattices offer the opportunity to realize a multitude of systems  
and to study their quantum phenomena \cite{Morsch,Lewenstein,Bloch,Lewenstein1,Yukalov}.
Moreover, recent experimental advances in optical trapping allow to control the size and atom number of these quantum 
systems, and furthermore include the tunability of the atomic   
interactions via Feshbach resonances \cite{Zurn,Serwane,Chin}. 
A promising research direction in this context is the non-equilibrium quantum dynamics for finite atomic ensembles.  
Here, the most frequently considered setting is a quantum quench (see Refs. \cite{Polkovnikov,Calabrese,Langen} and references therein),  
where one explores the quantum evolution 
after a sudden change of an intrinsic system parameter such as the interaction strength \cite{Kollath,Kollath1,Mistakidis,Mistakidis1}. 
A complicating feature of the non-equilibrium dynamics  
is the presence of interactions at a level that often precludes the use of a perturbative analysis and/or   
mean-field (MF) approximation. Specifically, the dynamics beyond the paradigm of linear response has been a 
subject of growing theoretical interest \cite{Fetter,Kinnunen,Proukakis,Alon_lr,Alon_lr1,Alon_lr2,Alon_lr3,Theisen,Beinke} triggered by the recent progress in ultracold atom 
experiments particularly in one spatial dimension \cite{Kinoshita,Kinoshita1,Haller,Paredes}. 

Referring to few-body systems in finite optical lattices, it has been shown \cite{Mistakidis,Mistakidis1} that following an interaction quench several tunneling pathways  
can be excited as well as collective behavior such as the cradle or breathing mode are observed.    
Furthermore, a sudden raise of the interactions \cite{Mistakidis} couples
one of the tunneling modes with the cradle mode giving rise to a resonant behavior.  
On the other hand, a sudden decrease of the inter-particle repulsion \cite{Mistakidis1} excites the 
cradle mode only for setups with a filling larger than unity and no mode coupling can be observed. 
From this it is evident that in order to steer the dynamics the considered quench protocol plays   
a key role. 
Naturally, one can then generalize the underlying protocol to a multiple interaction quench (MIQ) scenario, which consists of 
different sequences of single quenches. A specific case would be a quench followed by its 'inverse' namely by going back to the original interaction strength (single pulse).   
This enables the system to dynamically return to its original Hamiltonian within certain time intervals and the question emerges what properties 
induced by the quench persist during the longer time evolution.     
Very recently \cite{Chen} a study of the effects of the MIQ protocol on the one- and two-body correlation functions 
of a three-dimensional ultracold Bose gas has been performed using the time-dependent Bogoliubov approximation. 
It has been shown that the system produces more elementary excitations with 
increasing number of MIQs, while the correlation functions tend to a constant value for long evolution times. 

In the present work, we provide a multimode treatment of few bosons in finite optical lattices in one spatial dimension, where all correlations are taken into account.   
Such an approach is very appropriate in order to extract information on the resulting many-body dynamics 
and in order to obtain the complete excitation spectrum.  
This will allow us to explore how the MIQ protocol, reflected by the different temporal 
interaction intervals, affects the system dynamics and as a consequence the persistence of the emergent various collective modes during the evolution. 

Several protocols varying the number of quenches are hereby investigated.  
Our focus is on the regime of intermediate interaction strengths, where current state of the art analytical approaches are not applicable.  
The lowest-band tunneling dynamics involves several channels following a quench of increasing interaction, while only a few persist when the system is quenched back.  
Furthermore, the intrawell excited motion is described by the cradle and the breathing modes being initiated by the over-barrier transport which is a 
consequence of the quench to increased interactions. 
We find that in the course of the MIQ the cradle mode persists for all times,   
while the breathing mode possesses distinct frequencies depending on the different time intervals of the MIQ.  
In contrast to the single quench scenario \cite{Mistakidis,Mistakidis1} here by tuning the parameters of the MIQ 
we can manipulate both the interwell tunneling and the intrawell excited modes. 
Moreover, the higher-band excitation dynamics is explored in detail. 
A monotonic increase of the excited to higher-band fraction for larger quench amplitudes is observed and a non-linear dependence on the 
time interval of a single quench (pulse width) is revealed.     
Remarkably the interplay between the quench amplitude and the pulse width yields a tunability of  
the higher-band excitation dynamics.   
This observation indicates a substantial degree of controllability of the 
system under a MIQ protocol, being an important result of our work. 
Moreover, it is shown that in the course of a certain pulse the presence of increased interactions leads to a 
periodic population transfer between different lattice momenta, while for the time intervals of the initial interaction strength this does not happen. 
The frequency of the above-mentioned periodicity possesses a power-law dependence on the quench amplitude. 

This work is organized as follows. In Sec. \ref{protocol} we introduce the quench protocol and the multiband expansion as an analysis tool.  
Sec. \ref{dynamics} focuses on the detailed investigation of the impact of the  
MIQ on the quantum dynamics for filling factors larger than unity whereas  
Sec. \ref{multiwell} presents the dynamics for  
filling factors smaller than unity. We summarize our findings and present  
an outlook in Sec. \ref{conclusions}. Appendix A describes our computational method and delineates the convergence of our numerical results.  
\begin{figure}[ht]
       \centering
          \includegraphics[width=0.4\textwidth]{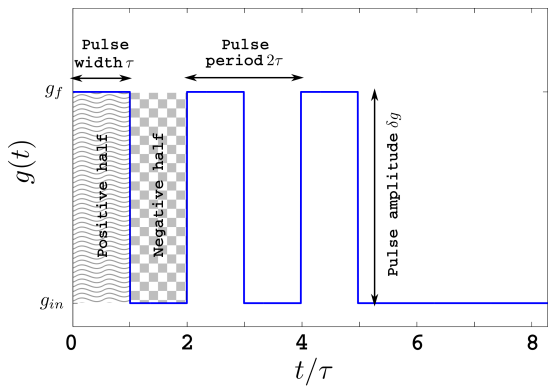}
               \caption{Sketch of a triple pulse MIQ protocol, $g(t)$, with pulse width $\tau$. $g_{in}$ ($g_{f}$) refer 
               to the pre-(post) quenched interaction strength and $\delta g=g_f-g_{in}$ is the pulse or quench amplitude.}
               \label{pulses}
\end{figure}

\section{Quench protocol and multiband expansion}
\label{protocol}

We consider $N$ identical bosons each of mass $M$ confined in an $m$-well optical lattice. 
The many-body Hamiltonian reads
\begin{equation} 
H=\sum_{i=1}^{N}\left(\frac{p_i^2}{2M}+V_{tr}(x_i)\right)+\sum_{i<j}V_{int}(x_i-x_j, \tau, n_p, t), \label{Hamilt}
\end{equation} 
where the one-body part of the Hamiltonian builds upon the one-dimensional lattice
potential $V_{tr}({x}) = {V_0}~{\sin ^2}(k{x})$. The latter is  
characterized by its depth ${V_0}$ and periodicity $l$, with $k = \pi
/l$ denoting the wave vector of the counter propagating lasers which form the optical
lattice. To restrict the infinitely extended trapping
potential $V_{tr}({x})$ to a finite one with $m$ wells and
length $L$, we impose hard wall boundary conditions at the
appropriate positions, $x_{m} = \pm
\frac{{m\pi }}{{2k}}$.  
Furthermore, $V_{int}(x_i-x_j,t, \tau, n_p)=g(t,\tau,n_p)\delta (x_i-x_j)$ 
corresponds to the contact interaction potential between particles located at positions $\{{x_i}\}$ with $i =1,2,...,N$. 

To trigger the dynamics we employ a MIQ protocol.       
At $t=0$ the interatomic interaction is quenched from the initial value $g_{in}$ to a final amplitude $g_f$,    
maintaining $g_{f}$ (positive half) for time $\tau$ (pulse width). Then, the interaction strength is quenched back from   
$g_f$ to its initial value $g_{in}$, maintaining this value $g_{in}$ (negative half) for time $\tau$. This procedure is repeated 
according to the number of the pulses $n_p$, see Fig. \ref{pulses} for the case of three pulses.   
Therefore, our protocol reads
\begin{equation} 
\begin{split}
\label{g_t}
g(t,\tau, n_p)=&g_{in}+(g_f-g_{in})\times\\&\sum_{i=0}^{n_p-1}\left[\Theta(t-2i\tau)\Theta((2i+1)\tau-t)\right].
\end{split}
\end{equation} 
Here, each pulse is modeled by a temporal step function depending on the parameters  
$n_p$ and $\tau$ which refer to the number of the considered pulses and the pulse width respectively.  
Moreover, $\delta g=g_f-g_{in}$ denotes the quench amplitude of the MIQ. 
Experimentally, the effective interaction strength in one dimension can be tuned either via the three-dimensional scattering length by using 
a Feshbach resonance \cite{Kohler,Chin} or by a change of the corresponding transversal 
confinement frequency $\omega_\bot$ \cite{Olshanii,Kim,Giannakeas}. 

For reasons of simplicity we rescale the Hamiltonian (\ref{Hamilt}) in units of the recoil 
energy ${E_R}=\frac{\hbar^2k^2}{2M}$. Thus the length, time and frequency scales are given in units of $k^{-1}$, $\omega_R^{-1}=\hbar E_R^{-1}$ 
and $\omega_R$ respectively. To include three localized single-particle Wannier states in each well 
we employ a sufficiently large lattice depth of $V_0=10.0 E_R$. Finally, for convenience we set $\hbar=M=k=1$. 
Hence, all quantities below are given in dimensionless units.

To solve the underlying many-body Schr{\"o}dinger equation we employ the Multi-Configuration Time-Dependent Hartree method for
Bosons (MCTDHB) \cite{Alon,Alon1}. 
In contrast to the MF approximation, 
within this method we take all correlations into account and employ a variable number of variationally optimized 
time-dependent single-particle functions (see Appendix A for more details). 
Below, when comparing with the MF approximation we will refer to MCTDHB as the correlated approach. 
For the interpretation and analysis of the induced dynamics it is preferable to rely on a time-independent 
many-body basis rather than the time-dependent one used for our numerical calculations. 
We therefore project the numerically obtained wavefunction on a time-independent 
number state basis consisting of single-particle states localized on each lattice site. 
Thus, the total wavefunction is expanded in terms of non-interacting multiband Wannier number states. 
The Wannier states between different wells possess a fairly small 
overlap for not too high energetic excitation as the employed lattice potential ($V_0=10.0 E_R$) is deep enough. 
Then, a many-body bosonic wavefunction for a system of $N$ bosons, $m$ wells 
and $j$ localized single-particle states \cite{Mistakidis,Mistakidis1} reads
\begin{equation}
\label{wannier} 
\ket{\Psi(t)}=\sum_{\vec{n}}C_{\vec{n}}(t)\ket{\vec{n}}
\end{equation} 
where $\ket{\vec{n}}\equiv\ket{\bigotimes_{\lambda=0}^{j-1}n_1^{(\lambda)},\bigotimes_{\lambda=0}^{j-1}n_2^{(\lambda)},...,\bigotimes_{\lambda=0}^{j-1}n_m^{(\lambda)}}$ denotes 
the multiband Wannier number state. 
Each element can be decomposed as $\bigotimes_{\lambda=0}^{j-1} n_i^{(\lambda)}=n_i^{(0)}\otimes n_i^{(1)}\otimes...\otimes n_i^{(j-1)}$, where  
$n_i^{(\lambda)}$ denotes the number of bosons being localized in the
$i$-th well, and $\lambda$-th band satisfying the closed subspace constraint
$\sum_{i=1}^m\sum_{\lambda=0}^{j-1} n_i^{(\lambda)} = N$.   
For instance, in a setup with $N=4$ bosons confined in a triple well  
$m=3$, being our workhorse in the following, which includes $\lambda=3$ single-particle states, the state 
$\ket{1^{(0)}, 1^{(0)}\otimes1^{(1)},1^{(0)}}$ 
indicates that in every well 
one boson occupies the zeroth excited band, but in the middle well there is one extra boson  
localized in the first excited band. 
For this setup we can identify four different energetic classes of number states.  
The single pairs $\{ {\left| {2^{(I_1)},1^{(I_2)},1^{(I_3)}}\right\rangle}+\circlearrowright\}$ (SP), 
the double pairs $\{ {\left| {2^{(I_1)},2^{(I_2)},0^{(I_3)}} \right\rangle}+\circlearrowright\}$ (DP), 
the triples $\{ {\left| {3^{(I_1)},1^{(I_2)},0^{(I_3)}} \right\rangle}+\circlearrowright\}$ (T), 
and the quadruples $\{ {\left| {4^{(I_1)},0^{(I_2)},0^{(I_3)}}\right\rangle}+\circlearrowright\}$ (Q), where $\circlearrowright$  
stands for all corresponding permutations and $\bold{I}=(I_1,I_2,I_3)$ indicates the order of the degree of excitation. 
For our purposes we only consider the corresponding subclass with isoenergetic 
states, e.g. for the double pairs $\{ {\left| {2^{(I_1)},2^{(I_2)},0^{(I_3)}}
\right\rangle}, {\left| {0^{(I_1)},2^{(I_2)},2^{(I_3)}} \right\rangle}, {\left| {2^{(I_1)},0^{(I_2)},2^{(I_3)}}\right\rangle}\}$.  
To characterize the eigenstates in terms of number states we adopt the compact notation $\ket{s}_{\alpha;\bold{I}}$, 
where $s$ denotes the spatial occupation and $\alpha$ relates to each of the above classes. 
For instance, $\{\ket{s}_{1;\bold{I}}\}$ with $\bold{I}=(1,1,0)$ represents 
$\{ {\left| {2^{(1)},1^{(1)},1^{(0)}} \right\rangle}$, ${\left| {2^{(1)},1^{(0)},1^{(1)}}
\right\rangle}$, ${\left| {1^{(0)},2^{(1)},1^{(1)}} \right\rangle}$, ${\left| {1^{(0)},1^{(1)},2^{(1)}}
\right\rangle}$, ${\left| {1^{(1)},1^{(0)},2^{(1)}} \right\rangle}$, ${\left| {1^{(1)},2^{(1)},1^{(0)}}
\right\rangle}\}$ and $s$ runs from 1 to 6.

\section{Quench dynamics for filling $\nu>1$}
\label{dynamics}

In this section the non-equilibrium dynamics following the MIQs for a system with filling factor $\nu>1$ is analyzed.    
The system is initially prepared in the ground state of four bosons confined in a triple well with interparticle repulsion 
$g_{in}=0.1$. It is thus dominated by the number state $\ket{1^{0},2^{0},1^{0}}$.   
To induce the dynamics we focus on a double and five pulse quench protocol [see Eq.(\ref{g_t}) for $n_p=2$ or $n_p=5$ and $\tau=50$ or $\tau=25$ respectively] 
and compare with the results for a single interaction quench. 

\begin{figure}[ht]
       \centering
          \includegraphics[width=0.5\textwidth]{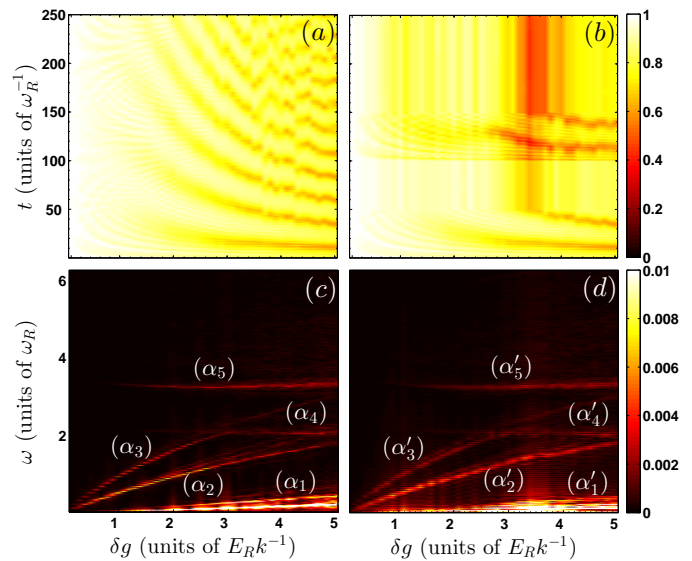}
               \caption{As a function of the quench amplitude $\delta g$ are shown: ($a$), ($b$) fidelity evolution for a single quench and a double pulse ($n_p=2$) MIQ  
               respectively and ($c$), ($d$) the corresponding fidelity spectra. Parameter values are $g_{in}=0.1$, $\tau=50$ and $N=4$.}
\label{fid_over_g}
\end{figure}

\subsection{Tunneling dynamics}

To investigate the dynamical response we employ the fidelity evolution $F(t;\tau)=\left|\braket{\Psi(0)|\Psi(t;\tau)}\right|^2$, being the 
overlap between the instantaneous and the initial wavefunction \cite{Venuti,Gorin,Campbell,March}.  
Following a single quench, see Fig. \ref{fid_over_g} ($a$), two different dynamical regions arise in the fidelity evolution.   
For $0.1\le g_f \le 1.0$ the system is only weakly perturbed since $F(t)\approx1$. 
For $g_f\ge 1.0$ the fidelity deviates significantly from unity and exhibits in time an oscillatory pattern.  
These oscillations are amplified with increasing quench amplitude and characterized both by a higher amplitude and frequency due to the increasing deposition of energy into the system. 
For the double pulse protocol the dynamical response is altered, as compared to the single quench scenario, and it is characterized by four distinct temporal regions, see Fig. \ref{fid_over_g} ($b$).  
For $t<\tau$ the same pattern as for the single quench is, of course, observed as the two protocols are identical within this time interval, i.e. $g(t<\tau)=g_f$.   
At $t=\tau$ the system is quenched back to $g_{in}$ and the oscillation of the fidelity almost vanishes.  
Then, $F(\tau<t<2\tau)\approx F(t=\tau)$ where the value $F(t=\tau)$ depends strongly on the phase of the oscillation at $t=\tau$ and therefore on $\delta g$.    
During the positive half of the second pulse $2\tau<t<3\tau$ an oscillatory pattern is observed, possessing the same frequencies with those occuring   
during the positive half of the first pulse. The system is driven further away from the initial state as more energy is added.  
Note that the dominant frequency of the oscillation depends on $\delta g$ as in the single quench scenario, see Fig. \ref{fid_over_g} ($a$).   
At $t=3\tau$ the system is quenched back to $g_{in}$ and the oscillatory behavior of the response again vanishes. 
Hereafter, $F(t>3\tau)\approx F(t=3\tau)$.  
Remarkably enough, for $3.2\le g_f \le 3.7$ the fidelity reduces significantly after the second pulse to the value 
$F(t>3\tau)=0.44$ at $g_f=3.4$. The existence of such strong response regions for certain combinations of $\delta g$ and $\tau$ 
is caused by the MIQ scenario and will be addressed below in more detail. 

To identify the corresponding tunneling modes that participate in the dynamics we inspect the spectrum of the fidelity \cite{Mistakidis,Mistakidis1,Mistakidis2} for the 
single quench [Fig. \ref{fid_over_g} ($c$)] and the double pulse [Fig. \ref{fid_over_g} ($d$)] protocols. 
Both scenarios excite the same frequency modes possessing though some differences, caused by the fact that in the finite time intervals that the system is quenched back  
within the double pulse protocol the response remains mainly stable. 
The observed modes triggered by a single (double-pulsed) quench can be energetically categorized as follows:  
$(\alpha_1)$ [($\alpha'_1$)] tunneling within the SP category, $(\alpha_2)$ [($\alpha'_2$)] tunneling between the SP and DP categories and 
$(\alpha_3)$ [($\alpha'_3$)] tunneling between the SP and T categories.  
The latter two processes are reminiscent of the atom pair tunneling which has been experimentally detected in driven optical lattices \cite{Folling,Chen1}. 
To gain more insight into the spectrum of the double pulse scenario we have splitted the evolution into the different temporal regions that the protocol 
imposes, i.e. $g=g_f$ or $g=g_{in}$. 
As nearly no oscillations occur in the negative halves of the double pulse ($\tau<t<2\tau$ and $t>3\tau$) all tunneling branches except $a_1$ are then suppressed.   
Note here that, in principle, for $g=g_{in}$ all branches possess very small and nearly equal frequencies 
which are resolvable in the case of a large enough $\tau$.   
However, for $t<\tau$ (positive half of the first pulse) and $2\tau<t<3\tau$ (positive half of the second pulse) the above-mentioned three modes occur, see also Fig. \ref{fid_over_g} ($d$). 
The latter enables us to dynamically manipulate or even switch on and off certain tunneling processes due to the presence or absence of increased interactions.  
Finally, we remark that the branches denoted e.g. by $(\alpha_4)$, $(\alpha_5)$ refer to higher-band excitations and will be addressed below.

\subsection{Dominant intrawell excitations: The cradle and the breathing modes}

\begin{figure}[ht]
       \centering
          \includegraphics[width=0.5\textwidth]{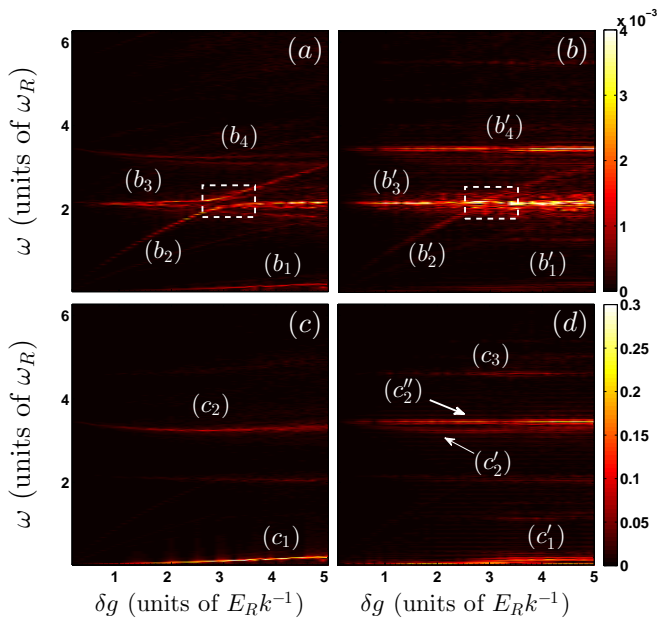}
               \caption{ As a function of the quench amplitude $\delta g$ are shown: ($a$), ($b$) spectrum of the intrawell asymmetry 
               $\Delta\rho_L(\omega)$ following a single quench and a double pulse ($n_p=2$) MIQ respectively.   
               Spectrum of $\sigma_M^2(\omega)$ following ($c$) a single quench and ($d$) a double pulse MIQ protocol. 
               Parameter values are $g_{in}=0.1$, $\tau=50$ and $N=4$.}
\label{crad_breath_A}
\end{figure}

Let us focus on the cradle and the breathing mode in the following.  
The cradle mode represents a dipole-like intrawell oscillation in the outer wells of the finite lattice. 
Following an interaction quench it is induced by an over-barrier transport of a boson initially residing in the central well (for a detailed description on the generation 
of this mode see \cite{Mistakidis,Mistakidis1}). 
It breaks the parity symmetry within the outer wells and can thus be quantified by the corresponding  
intrawell asymmetry of the wavefunction. For instance, in the left well $\Delta\rho_L(t)=\rho_{L,1}(t)-\rho_{L,2}(t)$, where $\rho_{L,1}(t)$ and $\rho_{L,2}(t)$ denote 
the spatially integrated densities of the left and the right half of the well. 
To investigate the frequencies that characterize the cradle mode and how they are influenced by the different quench protocols 
we employ the spectrum $\Delta\rho_L(\omega)=1/\pi \int dt e^{i\omega t}\Delta\rho_L(t)$. 
Previously \cite{Mistakidis} it has been shown that following a single interaction quench $\Delta\rho_L(\omega)$, as a function of the quench amplitude,  
possesses mainly two distinct frequency branches [see Fig. \ref{crad_breath_A} ($a$)]. 
The latter, refer to a tunneling mode $\ket{2^{(0)},1^{(0)},1^{(0)}} \rightleftharpoons \ket{3^{(0)},0^{(0)},1^{(0)}}$ 
[see branch $b_2$ in Fig. \ref{crad_breath_A} ($a$)] and 
an interband over-barrier process $\ket{1^{(0)},2^{(0)},1^{(0)}} \rightleftharpoons \ket{1^{(0)}\otimes 1^{(1)},1^{(0)},1^{(0)}}$ 
[see branch $b_3$ in Fig. \ref{crad_breath_A}($a$)] being identified as the cradle mode. 
Remarkably, these two modes come into resonance in a certain region of quench amplitudes [see the dashed rectangle in Fig. \ref{crad_breath_A} ($a$)], 
and therefore it is possible to couple the interwell (tunneling) with the intrawell (cradle) dynamics. 
However following a double pulse, see Fig. \ref{crad_breath_A}($b$), the aforementioned resonance is hardly visible as the tunneling mode 
is less pronounced compared to the single quench scenario [compare also Figs \ref{fid_over_g} ($c$), ($d$)].    
Indeed, the tunneling mode [see branch $b'_2$ in Fig. \ref{crad_breath_A} ($b$)] is present only when $g(t)=g_f$, while the cradle mode 
[see branch $b'_3$ in Fig. \ref{crad_breath_A} ($b$)] persists also after we quench back to $g_{in}$.  
The above can be explained as follows: when the interaction strength is reduced, the bosons do not possess the required energy to perform a second order tunneling 
process, and therefore the SP to T tunneling mode, see $b_2'$, is absent when we quench back.  
On the contrary, the cradle mode persists also when $g(t)=g_{in}$, $t>0$ as it is an intrawell mode and has already been initialized previously.  
Therefore, a tunneling process is required to initialize the cradle mode but is not a prerequisite for it to persist. 
As a consequence the coupling between the cradle mode and the SP to T tunneling mode disappears when $g=g_{in}$ and occurs only for $g=g_f$. 
Thus, using a MIQ protocol one can switch on and off the above described mode resonance [see also the dashed rectangle in Fig. \ref{crad_breath_A} ($b$)].   
Finally, we note that the energetically lower visible branch e.g. $b_1$ refers to tunneling within the SP mode [see also Fig. \ref{fid_over_g} ($c$)],   
while the energetically upper branch in both spectra located at $\omega \approx3.5$ belongs to the breathing mode and is explained below in detail. 

The breathing mode refers to an expansion and contraction of the bosonic cloud and can be 
excited by varying the interaction strength or the frequency of the trapping potential \cite{Abraham,Abraham1,Bauch}. 
Here, due to the lattice symmetry it is expected \cite{Mistakidis, Mistakidis1} to be more prone in the central well. 
To identify the breathing mode we employ the second moment $\sigma^2_M(t)=\bra{\Psi(t)}\hat{P}_M(x-X_{CM}^{(M)})^2\hat{P}_M\ket{\Psi(t)}$ 
within the spatial region of the middle well (denoted by the index $M$). Here, the operator $\hat{P}_M=\int^{\pi/2}_{-\pi/2}dx \ket{x}\bra{x}$ projects onto 
the spatial region of the middle well and 
$X_{CM}^{(M)}=\int_{-\pi/2}^{\pi/2}dx(x-x_0^M)\rho_M(x)/\int_{-\pi/2}^{\pi/2}dx\rho_M(x)$, $x_0^{M}$ and $\rho_M(x)$    
refer to the center of mass, the center position and    
the single-particle density of the middle well respectively. 
To investigate the frequency spectrum of the breathing mode we employ $\sigma^2_M(\omega)=1/\pi\int dt e^{i\omega t} \sigma^2_M(t)$ \cite{Klaiman,Klaiman1,Klaiman2}. 
For a single interaction quench, it has been shown \cite{Mistakidis} that $\sigma^2_M(\omega)$ possesses two distinct   
frequency branches, shown in Fig. \ref{crad_breath_A} ($c$). 
The upper branch (denoted as $c_2$) refers to the second order process 
$\ket{1^0,1^0,1^0\otimes 1^2}\rightleftharpoons \ket{1^0,2^0,1^0} \rightleftharpoons \ket{1^0\otimes 1^2,1^0,1^0}$, which indicates the presence of a 
global interwell breathing mode induced by the over-barrier transport.  
The fact that the breathing mode is also visible in the intrawell asymmetry spectrum [see Fig. \ref{crad_breath_A} ($a$)] of the left (right) well is another indication  
that it is indeed a global mode.  
The lower branch (denoted as $c_1$) corresponds to the interwell tunneling mode $\ket{1^0,2^0,1^0} \rightleftharpoons \ket{2^0,1^0,1^0}$. 
Both of the above branches weakly depend on the quench amplitude $\delta g$.   
Turning to the double pulse, [Fig. \ref{crad_breath_A} ($d$)] the above two branches now indicated by $c'_1$ and $c'_2$ persist   
but also two additional and $g_f$-independent branches marked by $c''_2$ and $c_3$ appear above $c'_2$. 
Note here that $\omega_{c_2''}=\lim_{g_f\to g_{in}}\omega_{c_2'}$ showing that these branches stem from the same eigenfrequencies, 
while the branch $c_3'$ refers to an admixture of higher-band states. 
Importantly, the $g_f$-independent branches exist only during the time intervals 
of $g_{in}$ i.e. for $\tau<t<2\tau$ and $t>3\tau$, whereas the $g_f$-dependence occurs only during the positive halves of the MIQ i.e. for $g=g_f$.   
To conclude, the double pulse MIQ protocol gives rise to two additional $g_f$-independent branches of breathing dynamics. 
The latter suggests that by tuning the intrinsic parameters of the MIQ protocol one can steer the induced breathing dynamics.

\subsection{Excitation dynamics}
\label{excite}

To gain a deeper understanding of the excitation dynamics, we investigate in the following the occupation of higher-band states during the time evolution.  
We consider the probability to find $N_0<N$ bosons in the $\lambda$-th band 
\begin{equation}
\begin{split}
\label{excit_prob}  
P_{N_0}^{(\lambda)}(t;\tau)=\sum_{\vec{n}\in\mathcal{N}_{N_0}^{(\lambda)}}|\braket{\vec{n}|\Psi(t;\tau)}|^2,
\end{split}
\end{equation}
where the notation $\vec{n}\in\mathcal{N}_{N_0}^{(\lambda)}$ denotes that the sum is performed over the configurations 
$\mathcal{N}_{N_0}^{(\lambda)} \equiv \{\vec{n}:\sum_{i=1}^{3}\sum_{\lambda=0}^{j-1} n_i^{(\lambda)}=4~~\text{and}~~\sum_{i=1}^3 n_i^{(\lambda)}=N_0\}$ that belong to the Hilbert space consisting of 
four particles from which $N_0$ reside in the $\lambda$-th band. 

The case of $\lambda=0$ and $N_0=N=4$ refers to the probability to find all four bosons within the ground band i.e. the energetically lowest band.  
Then, the above excitation probability reduces to  
$P_{N}^{(0)}(t;\tau)=\sum_{\vec{n}\in\mathcal{N}_{N}^{(0)}}|\braket{\vec{n}|\Psi(t;\tau)}|^2$. 
\begin{figure}[ht]
       \centering
          \includegraphics[width=0.45\textwidth]{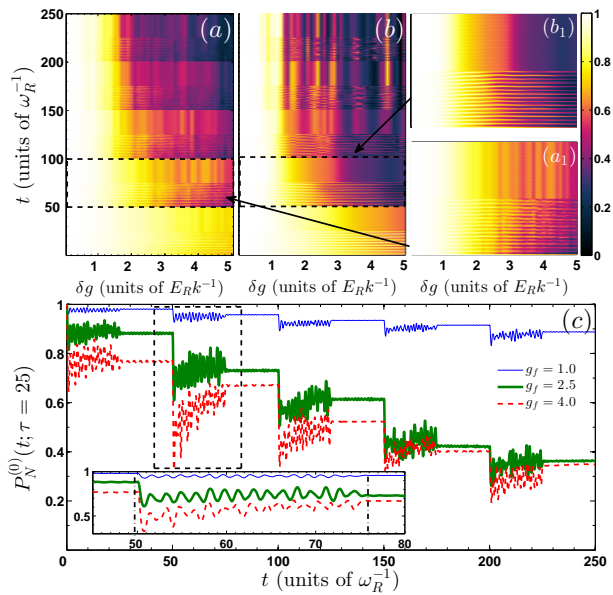}
               \caption{Time evolution of the probability $P_{N}^{(0)}(t;\tau=25)$ to find all four bosons within the ground band 
               with respect to the quench amplitude $\delta g$ following a five pulse ($n_p=5$) MIQ protocol.  
               ($a$), ($b$) $P_{N}^{(0)}(t;\tau=25)$ for varying $\delta g$ including correlations and for the MF approximation respectively. 
               The insets ($a_1$), ($b_1$) show $P_{N}^{(0)}(t;\tau=25)$ of ($a$), ($b$) respectively, only within the second pulse.    
               ($c$) Profiles of $P_{N}^{(0)}(t;\tau=25)$ for different $\delta g$ (see legend). 
               For better visibility of the oscillatory behavior during the positive half of the second pulse we show in the inset $P_{N}^{(0)}(45<t<80;\tau=25)$. 
               The system consists of four initially weakly interacting, $g_{in}=0.1$, bosons confined in a triple well. }
\label{ground_50_image}               
\end{figure}
To investigate the impact of the quench amplitude $\delta g$ we show in Fig. \ref{ground_50_image} ($a$) $P_{N}^{(0)}(t;\tau)$ following 
a five pulse MIQ protocol [see Eq.(\ref{g_t}) for $n_p=5$ and $\tau=25$].  
We observe that for $g_f\le1$ the occupations $P_{N}^{(0)}(t;\tau=25)$ are approximately unity and thus  
within this regime only to a minor degree excitations occur. 
However, for $g_f>1.0$ an oscillatory pattern in time is formed [see also Fig. \ref{fid_over_g} ($c$)], indicating the consecutive formation of higher-band excitations.   
In particular, within a positive half of the MIQ, i.e. $\ell\tau<t<(\ell+1)\tau$, $\ell=0,2,4,6,8$ large amplitude oscillations of $P_{N}^{(0)}(t;\tau=25)$ occur, while  
in the negative halves of the MIQ, i.e. $\ell\tau<t<(\ell+1)\tau$, $\ell=1,3,5,7$ and $t>9\tau$, the oscillatory behavior of $P_{N}^{(0)}(t;\tau=25)$ almost vanishes 
thereby forming an excitation plateau, see also Fig. \ref{ground_50_image} ($a_1$) which presents $P_N^{(0)} (2\tau<t<4\tau)$. 
In addition, focusing on $t>9\tau$ a nearly linear decrease of $P_{N}^{(0)}(t>9\tau;\tau=25)$ with increasing $\delta g$ is observed.  
This can be attributed to the fact that by using higher quench amplitudes we import more energy to the system and thus more excitations can be formed. 
Though some small deviations from this tendency exist, for instance we find a slightly lower $P_{N}^{(0)}(t;\tau)$ for $g_f=3.8$ than for $g_f=3.9$ 
[hardly visible in Fig. \ref{ground_50_image} ($a$)].  
To demonstrate the necessity of correlations for the description of the excitation dynamics we perform a comparison with the MF approximation.  
Fig. \ref{ground_50_image} ($b$) presents $P_{N}^{(0)}(t;\tau=25)$ within the MF approximation for varying $\delta g$.  
A similar qualitative overall behavior compared to the above analysis is observed.  
For $g_f<1$ the occupations $P_{N}^{(0)}(t;\tau=25)\approx1$, while for $g_f\geq 1$ $P_{N}^{(0)}(t;\tau=25)$ form oscillatory patterns 
within the positive halves of the MIQ and remain steady in the negative halves of the MIQ [see also Fig. \ref{ground_50_image} ($b_1$)].    
However, $P_{N}^{(0)}(t;\tau=25)$ is always lower when compared to the correlated approach, see Figs. \ref{ground_50_image} ($a$), ($b$) 
and in particular Figs. \ref{ground_50_image} ($a_1$), ($b_1$) which show $P_N^{(0)}(t;\tau)$ during the second pulse. 
Obviously, the oscillation amplitudes  
during the positive halves of the MIQ as well as the values of $P_N^{(0)}(t;\tau)$ for $g=g_{in}$ are larger within the MF approximation than the 
correlated approach.    
In addition, the linear dependence of $P_{N}^{(0)}(t>9\tau;\tau=25)$ is lost within the MF approximation 
and therefore we cannot observe an overall tendency of the excitation 
probability with increasing interparticle repulsion.  
To explicitly demonstrate the excitation process we show in Fig. \ref{ground_50_image} ($c$)    
various profiles of $P_{N}^{(0)}(t,\tau)$, taken from Fig. \ref{ground_50_image} ($a$), for different $\delta g$.  
For very short times $P_{N}^{(0)}(t,\tau)$ drops to a lower value and subsequently oscillates  
with an amplitude smaller than the initial decrease exhibiting multiple frequencies.   
Note that both the initial decrease as well as the amplitude and the oscillation frequency depend on $\delta g$, see also branch $c_2'$ in Fig. \ref{crad_breath_A} ($d$).  
During the positive halves of the MIQ , i.e. $\ell\tau<t<(\ell+1)\tau$, $\ell=0,2,4,6,8$, $P_{N}^{(0)}(t;\tau)$ 
oscillates with a decreasing amplitude (particularly for larger $\delta g$), but 
$\bar P_{N}^{(0)}(\tau)=1/T \int_{\ell\tau}^{(\ell+1)\tau} dt P_{N}^{(0)}(t;\tau)$ increases   
[e.g. see the dashed red line in the inset of Fig. \ref{ground_50_image} ($c$)].  
To gain more insight into the oscillation frequencies during the positive half of the MIQ protocol, we calculate the spectrum 
$P_N^{(0)}(\omega)=1/\pi\int_{0}^{T}dtP_{N}^{(0)}(t)e^{i\omega t}$. 
The latter shows two dominant branches from which the first one matches   
the frequency of the cradle mode [see branch $b'_3$ in Fig. \ref{crad_breath_A} ($b$)] and the other one corresponds to the frequency of the weakly  
$\delta g$-dependent breathing mode [see branch $c'_2$ in Fig. \ref{crad_breath_A} ($d$)]. 
At the end of the positive half, the amplitude of the above-mentioned oscillation suddenly decreases  
and $P_{N}^{(0)}(t;\tau)$ remains almost steady exhibiting only tiny oscillations (excitation plateaus).    
We again remark that the value of $P_{N}^{(0)}(t;\tau)$ in a negative half of a certain pulse, where the excitation plateaus appear, strongly 
depends on the phase of the oscillation at $t=\ell\tau$, $\ell=1,3,...,9$ (see also below). 
\begin{figure}[ht]
\vspace{-8pt}
       \centering
          \includegraphics[width=0.5\textwidth]{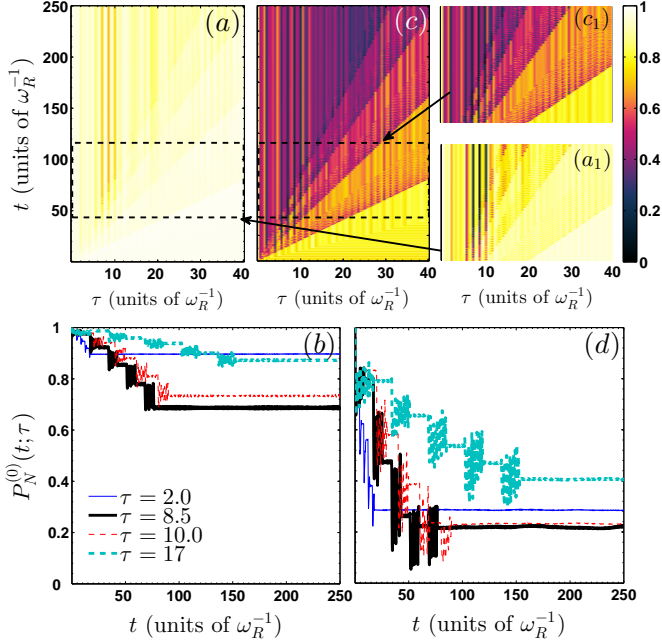}
               \caption{($a$) Time evolution of the probability to find all four bosons within the ground band $P_{N}^{(0)}(t;\tau)$ for varying  
               pulse width $\tau$ following a five pulse $n_p=5$ MIQ protocol with $\delta g=1.0$. ($b$) The corresponding profiles of $P_{N}^{(0)}(t;\tau)$ 
               for different values of $\tau$ (see legend). 
               ($c$), ($d$) The same as above but for $\delta g=2.9$. 
               The insets ($a_1$), ($c_1$) show $P_{N}^{(0)}(40<t<120;\tau=25)$ of ($a$), ($c$). The system consists of four initially weakly interacting $g_{in}=0.1$ bosons in a triple well. }
\label{ground_5p_image}               
\end{figure}

Next, we focus on the impact of the pulse width $\tau$ on the excitation dynamics. 
Fig. \ref{ground_5p_image} ($a$) shows $P_{N}^{(0)}(t;\tau)$ for varying pulse width and employing a five pulse MIQ protocol with $g_f=1.0$. 
Overall, we observe that $P_{N}^{(0)}(t;\tau)$ for fixed $\tau$ exhibits a similar oscillatory pattern as before within the positive halves of the MIQ and the formation of   
the excitation plateaus within the negative halves of the MIQ, see also Fig. \ref{ground_5p_image} ($a_1$).  
Also, $P_N^{(0)}(t;\tau)$ decreases with each additional pulse and remains almost steady after the last pulse.  
To illustrate the latter behavior, several profiles of $P_{N}^{(0)}(t;\tau)$ are shown in Fig. \ref{ground_5p_image} ($b$).   
In contrast to the approximately linear $\delta g$-dependence of $P_{N}^{(0)}(t>9\tau;\tau)$, we observe here that    
the fraction $1-P_N^{(0)}(t;\tau)$ of excitations depends on the pulse width in a non-linear manner, i.e. increasing $\tau$ does not necessarily lead to a smaller $P_{N}^{(0)}(t>9\tau;\tau)$.  
For instance, $P_{N}^{(0)}(t>9\tau,\tau=2.0)\approx0.89$, $P_{N}^{(0)}(t>9\tau,\tau=8.5)\approx0.68$ whereas $P_{N}^{(0)}(t>9\tau;\tau=10.0)\approx0.73$ [see Fig. \ref{ground_5p_image} ($b$)].   
It is also important to note that while $\delta g$ is the same for all pulses, the corresponding oscillation amplitude of $P_{N}^{(0)}(t;\tau)$ during a certain positive half of the MIQ     
is not fixed, indicating that it is not only affected by the quench amplitude but also depends on the pulse width.  
To further elaborate on the effects of the combination of $\delta g$ and $\tau$, Fig. \ref{ground_5p_image} ($c$) 
shows $P_{N}^{(0)}(t;\tau)$ for $g_f=2.9$ and varying $\tau$.  
As the quench amplitude is increased the system produces more excitations and after the pulses $\bar P_{N}^{(0)}(\tau)$ is much lower than in the case of 
$g_f=1.0$ [compare Figs. \ref{ground_5p_image} ($a_1$) and ($c_1$)].  
Overall, $P_{N}^{(0)}(t;\tau)$ behaves similar as in the case of $g_f=1.0$, see also the corresponding profiles in Fig. \ref{ground_5p_image} ($d$), 
and the non-linear dependence on the pulse width is again present.  
The oscillation amplitudes of $P_{N}^{(0)}(t;\tau)$ within the positive halves of the MIQ protocol are larger as compared to the case of $g_f=1.0$. This $\delta g$-dependence of the oscillation amplitude  
has been observed also for other quench amplitudes (results not shown here).   
Finally, $P_{N}^{(0)}(t>9\tau;\tau)$ remains almost steady, while a larger $\delta g$ leads in general to a lower $P_{N}^{(0)}(t>9\tau;\tau)$. 
However, exceptions do in principle exist indicating the significance of the optimal combination of $\delta g$ and $\tau$. 
\begin{figure*}[ht]
        \centering
           \includegraphics[width=0.8\textwidth]{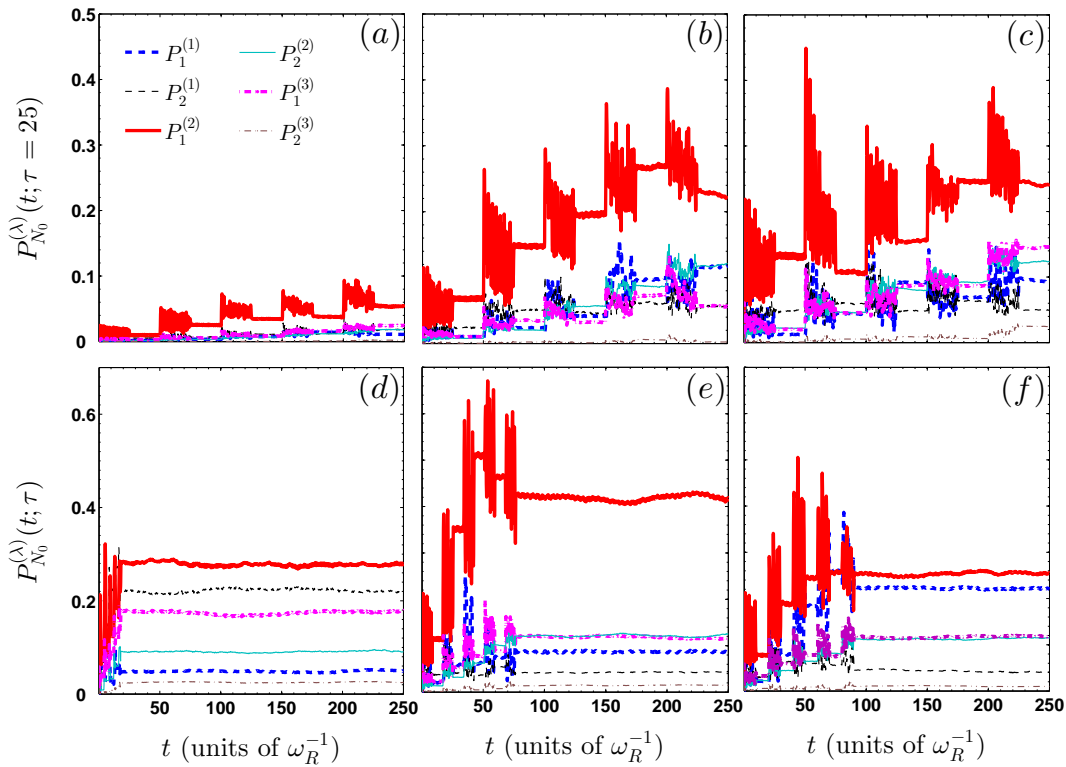}
                \caption{Time evolution of the probability to find one or two bosons in higher-bands (see legend) following a five pulse MIQ protocol with fixed 
                pulse width $\tau=25.0$ and varying amplitude 
                ($a$) $\delta g=1.0$, ($b$) $\delta g=2.5$ and ($c$) $\delta g=4.0$. ($d$), ($e$), ($f$) The same as above but for fixed quench amplitude $\delta g=4.0$ and varying pulse width   
                ($d$) $\tau=2$, ($e$) $\tau=8.5$ and ($f$) $\tau=10$. The system consists of four bosons initially prepared in the ground state with $g_{in}=0.1$ of a triple well. }
 \label{ex_5p_plot}
\end{figure*}

To gain a deeper understanding of the underlying excitation processes during the evolution we explore the probability of 
finding $N_0<N$ bosons within the $\lambda$-th band, see Eq. \ref{excit_prob}.   
For instance, 
$P_{N_0}^{(\lambda)}(t;\tau)=\sum_{\vec{n}\in\mathcal{N}_{N_0}^{(\lambda)}}|\braket{\vec{n}|\Psi(t;\tau)}|^2$, 
for $\lambda=1,2$ represent the probability to find $N_0$ bosons 
within the first or second excited band respectively.   
More precisely, below, we investigate the probability to have one or two bosons in the first, second or third excited band as higher-lying states are 
not significantly occupied in our system. 
First we shall study the effect of $\delta g$ on the different excitation processes by considering a five pulse MIQ protocol with fixed pulse width $\tau=25$ and varying $\delta g$.  
Figs. \ref{ex_5p_plot} ($a$)-($c$) show the probability to find one or two bosons in the first, second or third excited band. 
For all shown cases, the same overall excitation pattern [e.g. see $P_{N}^{(0)}(t;\tau)$ in Fig. \ref{ground_50_image} ($c$)] is observed. Within a positive half of the 
MIQ $P_{N_0}^{(\lambda)}(t)$ oscillates, 
while in the negative halves of the MIQ it remains almost steady (excitation plateaus) possessing only tiny amplitude oscillations.  
Each pulse increases the value of $P_{N_0}^{(\lambda)}(t)$ and larger values of $\delta g$ lead to larger excitation probabilities as more energy is added to the system.  
Overall, we observe that a single-particle excitation to the second excited band $P_1^{(2)}(t;\tau)$, which refers to the breathing mode, possesses the main contribution. However, for 
increasing $\delta g$ also other and mainly higher excitation processes start to play a role and contribute significantly to the dynamics as shown in Figs. \ref{ex_5p_plot} ($b$), ($c$).  
These higher order excitations correspond to single and two-particle excitations in the first, second and third excited band possessing comparable 
amplitudes. Note that excitations higher than a two-particle excitation to the third excited band are negligible. 
\begin{figure}[ht]
       \centering
          \includegraphics[width=0.50\textwidth]{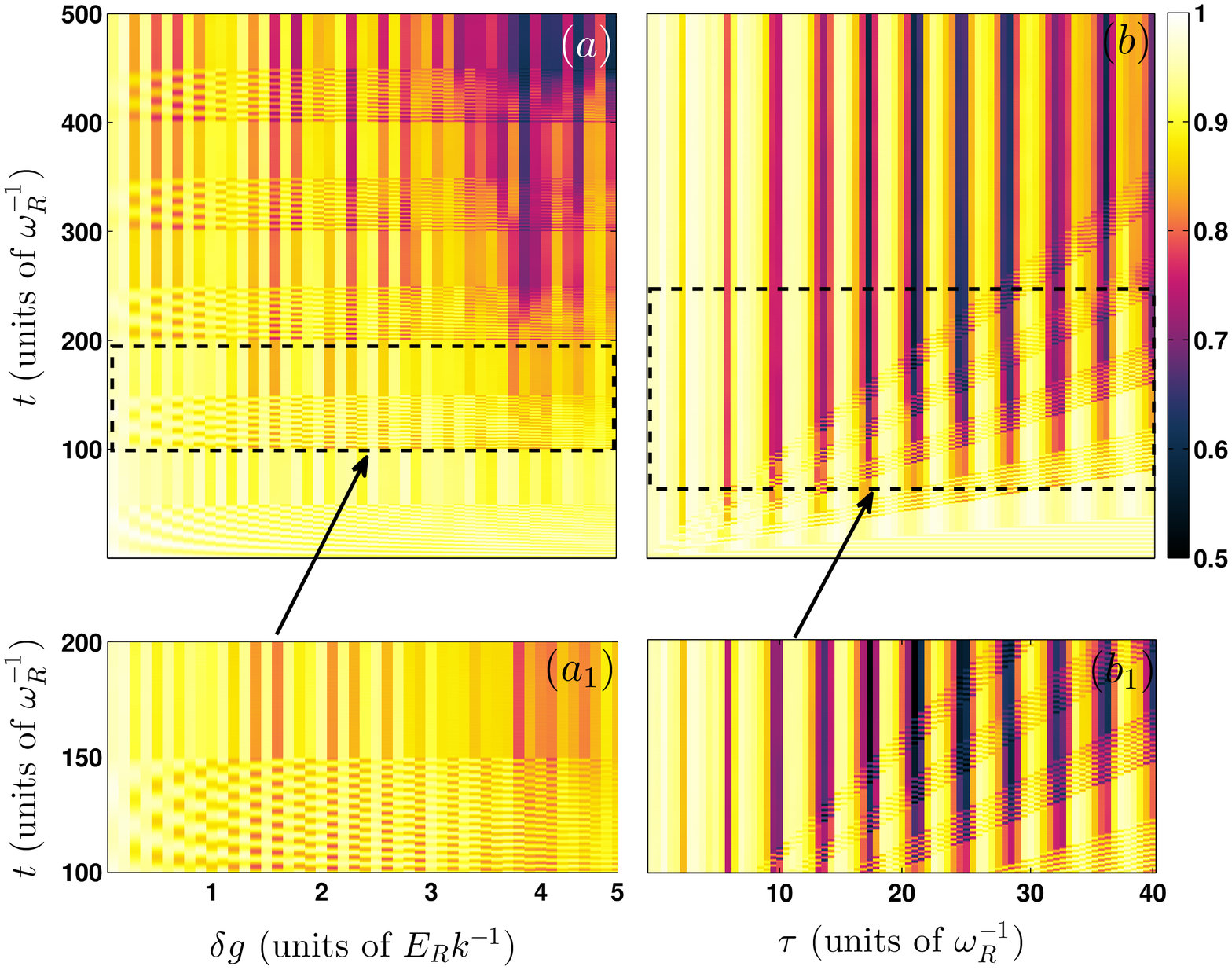}
               \caption{($a$) Fidelity evolution with varying $\delta g$, employing a five pulse $n_p=5$ MIQ protocol with $\tau=50$. 
               The inset ($a_1$) depicts $F(t;\tau=50)$ of ($a$) only within the duration of the second pulse.
               ($b$) The same as in ($a$) but with varying pulse width $\tau$ and $\delta g=2.9$. 
               The inset ($c_1$) shows $F(80<t<250;\tau)$ of ($c$).  
               The system consists of three bosons confined in an eight well potential.}
\label{fid_8well}
\end{figure}

Next, let us inspect the role of $\tau$ on $P_{N_0}^{(\lambda)}(t;\tau)$ employing the five pulse MIQ protocol with $g_f=4.0$. 
For $\tau=2.0$, see Fig. \ref{ex_5p_plot} ($d$), we observe a competition between  
$P_1^{(2)}(t;\tau)$, $P_2^{(1)}(t;\tau)$ and $P_1^{(3)}(t;\tau)$ possessing also  
the highest contributions after the pulses, namely $P_1^{(2)}(t>9\tau) \approx 0.28$, $P_2^{(1)}(t>9\tau) \approx 0.23$ and $P_1^{(3)}(t>9\tau) \approx 0.18$ respectively.   
$P_2^{(2)}(t;\tau)$ is to a lesser extent contributing with an amplitude $P_2^{(2)}(t>9\tau)\leq 0.1$. 
All the other excitations are significantly below $P_2^{(2)}(t;\tau)$. 
The final state after the last pulse, $t>9\tau$, exhibits many different excited modes.  
For $\tau=8.5$, see Fig. \ref{ex_5p_plot} ($e$), $P_1^{(2)}(t;\tau)$ clearly possesses the dominant contribution after the last pulse with $P_1^{(2)}(t>9\tau) \approx 0.42$. 
In addition, $P_2^{(2)}(t;\tau)$ and $P_1^{(3)}(t;\tau)$ are significantly smaller with comparable contributions around $0.1$.   
All the remaining states are negligible and their contributions are below $0.1$. 
Therefore, the parameter values $\tau=8.5$ and $g_f=4.0$ appear to be a good combination in order to achieve a single-particle excitation to the second excited band. 
For $\tau=10.0$, see Fig. \ref{ex_5p_plot} ($f$), $P_1^{(1)}(t;\tau)$ that mainly refers to the cradle mode and $P_1^{(2)}(t;\tau)$ are the dominant contributions  
with $P_1^{(1)}(t>9\tau) \approx 0.22$ and $P_1^{(2)}(t>9\tau) \approx 0.25$ respectively. 
A less dominant interplay is observed for the states $P_2^{(2)}(t;\tau)$ and $P_1^{(3)}(t;\tau)$ which fluctuate around $0.15$. 
The remaining excitation processes, e.g. $P_2^{(1)}(t;\tau)$ contribute below $0.05$.  
In this case we observe that the final state includes single-particle excitations to the first, second and third excited bands as well as a two-particle excitation 
to the second band. Other excitation processes are distributed below $0.05$ and do not contribute essentially to the final state.  
The above discussion suggests that for a fixed $\delta g$ (pulse width $\tau$) different excited states can be targeted by employing different pulse widths $\tau$ (quench amplitudes $\delta g$).   
Therefore, one can achieve a particular band occupation by choosing a specific combination of $\tau$ and $\delta g$. 
We remark that this picture is confirmed by employing different combinations of the number of pulses, pulse widths and final interaction strengths. 
\begin{figure}[ht]
       \centering
          \includegraphics[width=0.50\textwidth]{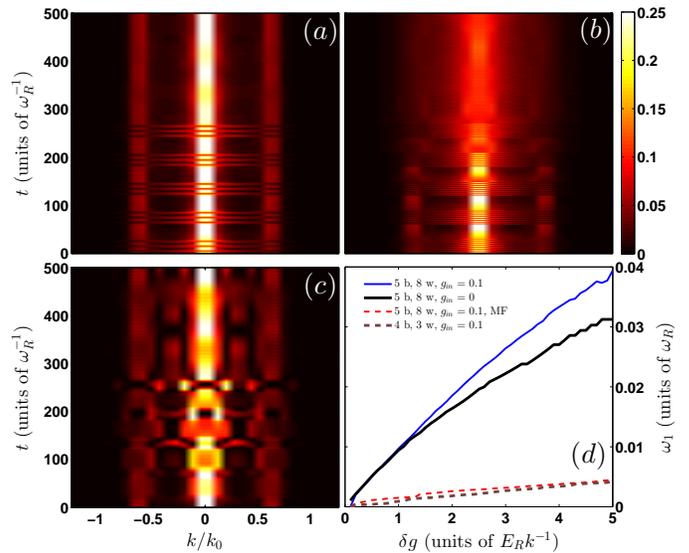}
               \caption{Momentum distribution as a function of time for ($a$) $\delta g=1.0$ and ($b$) $\delta g=2.9$. 
               ($c$) The same as ($a$) but within the MF approximation. The horizontal axis represents the lattice momenta 
               in units of the inverse lattice vector $k_0=\pi/l$. 
               The system consists of an eight well lattice potential with five bosons being subjected to a five pulse ($n_p=5$) MIQ characterized by $\tau=25$.  
               ($d$) Dominant oscillation frequency $\omega_1$    
               that appears in the momentum distribution for varying quench amplitude $\delta g$. Different curves correspond to different initial conditions,  
               approximations and system size (see legend).}
\label{momenta_8well}
\end{figure}

To demonstrate the applicability of our results for larger systems, in the following section, we proceed to the investigation of a system with
filling factor $\nu  < 1$.   
In particular, we shall show that the character of the excitation dynamics induced by a MIQ exhibits  
similar characteristics to the triple well case.

\section{Dynamics for fillings $\nu<1$}
\label{multiwell}

Let us now focus on a setup of three bosons confined in a lattice potential consisting of eight wells. 
For the ground state with filling factor $\nu<1$ a spatial redistribution 
of the atoms occurs with increasing interaction strength, i.e. the atoms are pushed from the central to 
the outer wells \cite{Brouzos}.  
Here, the initial state is the ground state for $g=0.1$, where 
the particles are predominantly localized in the center of the multi-well trap. 
Then, the ground state is dominated by Wannier number states of the 
form $|0,0,0,1, 2 ,0,0,0 \rangle$, $|0,0,0,1, 1 ,1,0,0 \rangle$, $|0,0,0,0,3,0,0,0 \rangle$ and their corresponding  
parity symmetric states, e.g. $|0,0,0,2,1,0,0,0 \rangle$,  due to the underlying spatial symmetry of the system. 

To induce the dynamics a five pulse MIQ protocol with $\tau=50$ is applied for $t>0$.  
Fig. \ref{fid_8well} ($a$) presents the fidelity evolution $F(t;\tau)$ for varying quench amplitude $\delta g$.  
The overall dynamical behavior is similar to the triple well case (see Sec. \ref{dynamics}).  
Indeed, during the positive halves of the MIQ the system is driven far from its initial ground state, while when it is quenched back it tends to a steady state. 
Each additional pulse, drives the system further away from its initial state. 
To visualize better the response of the system during a certain pulse Fig. \ref{fid_8well} ($a_1$) illustrates $F(2\tau<t<4\tau;\tau=50)$ i.e. during the second pulse as 
a function of $\delta g$. 
The fidelity shows an oscillatory pattern during the positive half and an almost fixed value in the corresponding negative half. 
Note here that the oscillatory pattern of $F(t;\tau)$ possesses multiple frequencies which mainly correspond to the different tunneling modes triggered by the MIQ. 
These frequencies become larger for increasing $\delta g$.    
Let us next examine the impact of the pulse width $\tau$ on the response of the system, namely we consider a five pulse MIQ with fixed $g_f=2.9$ and vary the 
pulse width, see Fig. \ref{fid_8well} ($b$).   
The dynamical response of the system resembles that for the triple well case (see Sec. \ref{dynamics}). It exhibits an oscillatory pattern within the positive 
halves of the MIQ protocol, tends to a steady state when 
$g(t)=g_{in}$ and increases with each additional pulse. 
The above description is illustrated in a transparent way in Fig. \ref{fid_8well} ($b_1$) where $F(80<t<250;\tau)$ is shown. 
Note here that $F(t;\tau)$ exhibits multiple frequencies during the evolution that refer to the induced tunneling dynamics.   
Finally, $F(t>9\tau;\tau)$ shows a non-linear dependence on $\tau$. 

To understand, whether signatures of parametric amplification of matter waves can be observed during the evolution we inspect 
the momentum distribution 
\begin{equation}
\label{momentum_distribution}
n(k,t)=\frac{1}{2\pi}\int \int dx dx' \rho_1(x,x',t)e^{-ik(x-x')t},
\end{equation}
where $\rho_1(x,x',t)$ denotes the one-body reduced density matrix, being obtained by tracing out all the bosons 
but one in the $N$-body wavefunction.  
We remark that the momentum distribution can be observed experimentally as it is accessible via time-of-flight measurements \cite{Bloch,Will,Bucker}.  
Fig. \ref{momenta_8well} ($a$) presents the time evolution of the 
momentum distribution for an eight well lattice potential with five bosons that are subjected to a MIQ of small quench amplitude and pulse width, namely $\delta g=1.0$ and $\tau=25$.     
As shown, employing a MIQ protocol the momentum distribution exhibits in time a periodically modulated pattern when $g=g_f$, e.g. see Fig. \ref{momenta_8well} ($a$) for $0<t<25$.  
Indeed, within the positive half of the MIQ $n(k,t)$ oscillates with frequency $\omega_1$ between the momenta $k_0=0$, $\pm k_0/2 \approx \pm 1.57$ i.e. it is  gradually transformed 
from a side peak structure (peaks at $k_0=0$, $\pm k_0/2$) to a broad maximum around $k_0=0$. 
On the contrary, in the negative halves of the MIQ ($\ell\tau<t<(\ell+1)\tau$, $\ell=1,3,5,7,9$) as well as after the last pulse $t>10\tau$ this side peak structure is preserved. 
Note here that the frequency $\omega_1$ does not depend on the considered pulse width. 
However, one can tune the time intervals of the periodic modulation by considering pulses with different $\tau$'s. 
The observed periodic population transfer between the $k_0=0$, $\pm k_0/2$ momenta is reminiscent of the parametric amplification of matter-waves.  
Similar observations have been made experimentally in different setups in Refs. \cite{Gemelke,Will,Kinoshita1}.  
This might pave the way for a more elaborated study of this process in the future, also for higher particle numbers and lattice potentials, 
but it is clearly beyond the scope of this work.  
We remark here that in the case of a single interaction quench only the above-mentioned periodic modulation within the positive half of the MIQ protocol can be achieved. 
Furthermore, the momentum distribution for a stronger interparticle repulsion i.e. $g_f= 2.9$, shown in Fig. \ref{momenta_8well} ($b$), 
exhibits a similar structure as above but with an increasing frequency within the positive halves of the MIQ.  
However, for large evolution times this behavior  
is blurred as an effect of the strong interaction which decreases the degree of coherence. 
Comparing with the MF calculation, shown in Fig. \ref{momenta_8well} ($c$), we observe that the periodic modulation of the populated lattice momenta 
within the positive halves of the MIQ is essentially lost and no clear signature of the effect of a pulse can be seen in $n(k,t)$.  
However, the activation of the additional lattice momenta is present but not in a systematic manner. 
This indicates the inescapable necessity of taking into account correlations for the description of the out-of-equilibrium dynamics. 
We remark here that in the case of larger filling factors where the presence of interparticle correlations is more dominant the failure of the MF approximation to 
capture certain features of the dynamics is even more prominent (results not shown here for brevity). 
Summarizing, the coherent MIQ dynamics leads to a periodic population transfer between different lattice momenta within a positive half of the MIQ and 
a side peak structure when the system is quenched back.  

To shed further light on the possible control of the dynamics we finally examine the dependence of the frequency $\omega_1$ of the periodic modulation during 
a positive half of the MIQ on several system parameters. 
Fig. \ref{momenta_8well} ($d$) presents $\omega_1$ with varying quench amplitude $\delta g$.  
As shown, $\omega_1$ depends strongly on the interparticle repulsion and in particular for increasing $\delta g$   
it possesses a power law behavior, namely 
\begin{equation}
\omega_1(g_f;N,g_{in})=\alpha g_f^b+c,
\end{equation} 
where $\alpha$, $b$, $c$ are positive constants. 
This is also in line with our previous observations on the evolution of the momentum distribution, see Figs. \ref{momenta_8well} ($a$), ($b$).  
Although, the periodic modulation of lattice momenta within the positive halves of the MIQ protocol is essentially lost within the corresponding MF approximation we 
also present the dominant frequency of $n(k,t)$ as a function of $\delta g$ in Fig. \ref{momenta_8well} ($d$). 
The obtained frequency dependence retains the above-mentioned power-law behavior but the corresponding frequencies for fixed $\delta g$ are smaller even 
for low quench amplitudes.  
The latter, is another manifestation of the failure of the single orbital approximation to accurately describe the induced dynamics. 
An additional intriguing question is whether $\omega_1$ depends on the initial interparticle repulsion and therefore possesses a many-body nature. 
Starting from a broader initial wavepacket, i.e. $g_{in}=0$ shown in the same Figure, $\omega_1$ is lower especially for $1.5<g_f<5$ where the system is significantly 
perturbed from its initial state [see also $\bar{F}(\tau)=1/T\int_0^T dt F(t;\tau)<0.8$ in Fig. \ref{fid_8well} ($a$)].      
To show that the general trend of $\omega_1$ is valid also for other system sizes, Fig. \ref{momenta_8well} ($d$) illustrates the obtained $\delta g$-dependence 
for four bosons confined in a triple well considering the same values for the initial system parameters [see also Fig. \ref{ground_50_image} ($a$)].    
Indeed, a similar functional form is observed, while the frequencies $\omega_1$ for the same $\delta g$ are reduced when compared to the eight well case.

\section{Conclusions and Outlook}
\label{conclusions}

We have explored the non-equilibrium quantum dynamics of multiply interaction quenched few boson ensembles confined in a finite optical lattice.   
Initially the system is within the weak interaction regime and sequences of interaction quenches to strong interactions and back are performed.    
To characterize the impact of the multiple pulses we study the interplay 
between the quench amplitude and the pulse width during the evolution. 
A variety of lowest-band interwell tunneling modes, a cradle mode and different breathing modes are excited.   
Focusing on the different time intervals of the MIQ protocol we identify the  
frequency branch of each process and the time intervals for which they exist.    
To further illustrate the peculiarity of a MIQ protocol we compare with the single quench scenario. 
We have analyzed the dynamical behavior by applying 
multi-band Wannier number states and identifyied for each of the above-mentioned processes the transitions  
between the dominant number states. 

The lowest-band interwell tunneling dynamics consists of three different energy channels which exist in the positive halves of the MIQ.  
When the system is quenched back only one tunneling mode survives.  
This raises the possibility to manipulate the tunneling dynamics within the different time intervals of the MIQ protocol. 
For instance, using different pulse widths we can switch on and off for chosen time intervals certain tunneling modes of the system.  

We then turned to the excited modes, i.e. the cradle and the breathing modes. 
The cradle mode 'ignores' the multipulse nature of the quench protocol  
and persists during the time evolution. 
However, the breathing mode shows a strong dependence on the instantaneous interatomic repulsion.  
Indeed, within the positive halves of the MIQ it possesses an interaction dependent frequency branch. 
However, in the negative halves of the MIQ the latter branch disappears and two new frequency branches appear which are interaction independent. 
As a result the system turns from the $\delta g$-dependent to the $\delta g$-independent branch providing further controllability.   
Furthermore, the excitation dynamics is investigated in detail. 
To analyze the dependence on the quench amplitude we focus on a fixed pulse width and vary the final interaction strength.  
It is shown that the excitation dynamics possesses a linear dependence on the quench amplitude, i.e. for increasing amplitude of the quench the 
amount of excitations as seen in the fidelity increase.  
For the dependence of the excitation dynamics on the pulse width we observe   
a non-linear dependence, i.e. there is no monotonic behavior of the produced excitations with varying pulse width. 
The latter implies that in order to control the excitation dynamics one has to use an optimal combination of the quench amplitude and the pulse width.   
Another prominent signature of the impact of the quenches is revealed by inspecting the momentum distribution. 
A periodic population transfer of lattice momenta within the positive halves of the MIQ protocol and a transition to a side peak structure in the negative halves of the MIQ are observed. 
This periodic population transfer of lattice momenta constitutes an independent signature of the excited energy channels within the positive halves of the MIQ protocol, allowing to study it from another 
perspective and to potentially measure it in corresponding experiments. 

Let us comment on a possible experimental realization of our setup.  
In a corresponding experiment weakly interacting bosons should be trapped in a one-dimensional optical superlattice being formed by two retroreflected laser beams. 
To form each supercell of the superlattice the first beam possesses a large wavenumber and intensity when compared to the second beam which forms each cell of the supercell. 
The above mentioned wavenumbers should be commensurate.
Then, the potential landscape of each supercell is similar to the one considered in the present study.    
Such an experimental implementation may be achieved either by the use of holographic masks \cite{Sherson} or by the 
modulation of the wavenumber \cite{Tli}. 
The interatomic repulsion can be tuned with the aid of a magnetic Feshbach resonance.   
The corresponding dynamical properties can then be probed with the recently developed single-site resolved imaging 
techniques (quantum microscope) \cite{Preiss,Fukuhara,Miranda}. 
We also remark that double occupancies can also be identified via Feshbach molecule formation \cite{Nagerl1,Winkler,Danzl}, while triple occupancies 
can be measured by inducing three-body recombination \cite{Esry,Kraemer,Knoop}.  

Finally, we provide an outlook onto possible future investigations. 
The achieved understanding of the non-equilibrium dynamics induced by multiple pulses of the interatomic repulsion  
may inspire similar investigations in other more complicated systems.  
A possible direction would be to apply our protocol to repulsively interacting dipolar systems and/or to include modulations of the lattice geometry.  
Certainly for larger particle numbers and sizes the question whether thermalization \cite{Srednicki,Rigol} occurs for 
long evolution times after the system has been quenched to its initial Hamiltonian is an intriguing one.

\appendix
\section{The Computational Method MCTDHB} \label{subsec:MCTDH}

To solve the time-dependent many-body Schr\"{o}dinger equation $\left( {i\hbar {\partial _t} - H} \right)\Psi (x,t) = 0$ we apply the 
Multi-Configuration Time-Dependent Hartree method for bosons \cite{Alon,Alon1,Streltsov} (MCTDHB). This method has been  
used extensively in the literature to explore the bosonic quantum dynamics, see for instance \cite{Streltsov,Streltsov1,Alon2,Alon3,Koutentakis,Mistakidis4}.  
The key idea of MCTDHB is to exploit time-dependently variationally optimized single-particle functions (SPFs) to form many-body states and thus to achieve an   
optimal truncation of the Hilbert space.  
The ansatz for the many-body wavefunction is taken as a linear combination of time-dependent permanents $|\vec{n} (t) \rangle$ with time-dependent  
weights $A_{\vec{n}}(t)$.  
Each time-dependent permanent $| \vec{n} (t) \rangle$ corresponds to a certain configuration of bosons that occupy $M$ variationally optimized     
SPFs $\left| \phi_j (t) \right\rangle$. 
In turn, the SPFs are expanded using a primitive time-independent basis $\{\ket{\chi}\}$ of dimension $M_{pr}$. 
The time evolution of the $N$-body wavefunction for the Hamiltonian under consideration reduces to the determination of 
the $A$-vector coefficients and the SPFs which obey the variationally obtained MCTDHB equations of motion \cite{Alon,Alon1,Streltsov}. 
We also remark that in the limiting case of $M=1$, MCTDHB reduces to the time-dependent Gross Pitaevski equation.  
\begin{figure}[ht]
       \centering
          \includegraphics[width=0.45\textwidth]{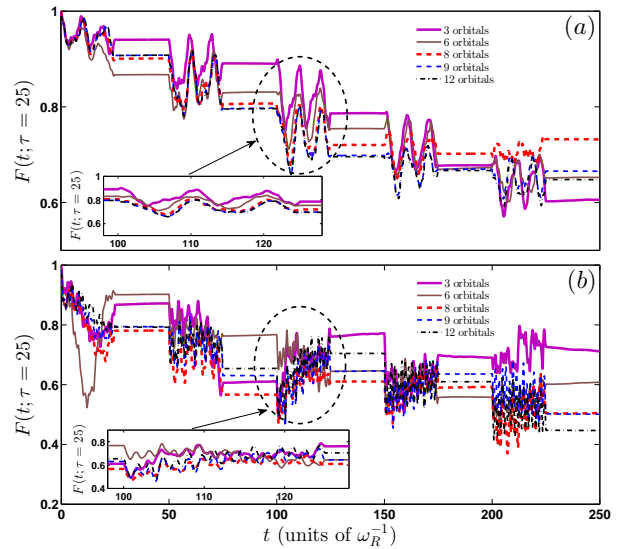}
               \caption{$F(t;\tau=25)$ for a varying number of SPFs (see legend) following a five pulse ($n_p=5$) MIQ with ($a$) $\delta g=1$ and ($b$) $\delta g=3.0$. 
               For better visibility of the evolution within the positive halves of the MIQ protocol we show in the insets $F(t;\tau=25)$ only during the third pulse. }
\label{convergence}
\end{figure}

For our implementation we have used a sine discrete variable
representation as a primitive basis for the SPFs. 
To prepare the system in the ground state of the Hamiltonian $H$, we rely on the relaxation method. The key idea
is to iteratively propagate some initial ansatz wavefunction $\left|\Psi ^{(0)} \right\rangle$ in imaginary time. 
This exponentially damps out all contributions
but the one stemming from the ground state like $\sim {e^{ - (E_m - E_0)\tau }}$ and therefore the system relaxes to the ground state 
(within the prescribed accuracy) after a finite time. 
To study the dynamics, we propagate the wavefunction by utilizing the appropriate Hamiltonian within the MCTDHB equations of motion.
Finally, let us remark that for our implementation we employed the Multi-Layer Multi-Configuration Hartree method for 
bosons \cite{Cao,Kronke} (ML-MCTDHB), which reduces to MCTDHB for the case of a single bosonic species as considered here.

To maintain the accurate performance of the numerical integration of the MCTDHB equations of motions we ensured that   
$\left| \langle \Psi(t) | \Psi(t) \rangle - 1 \right| < 10^{-8}$ and   
$\left| \langle \varphi_i(t) | \varphi_j(t) \rangle - \delta_{ij} \right| < 10^{-9}$ for the total wavefunction and the SPFs respectively. 
To conclude about the convergence of our simulations, we increase the number of SPFs 
and primitive basis states, thus observing a systematic convergence of our
results. In particular, we have used $M_{pr}=300$, $M=9$ ($M_{pr}=600$, $M=8$) for the triple well (eight wells).  
To be more concrete, in the following we shall briefly demonstrate the convergence procedure for the triple well simulations with increasing number of SPFs.
Figs. \ref{convergence} ($a$), ($b$) present $F(t;\tau=25)$ for different number of SPFs following a five pulse MIQ with small ($\delta g=1$) and large ($\delta g=3$) quench amplitudes respectively.  
In both cases a systematic convergence of the fidelity evolution (for $M>8$) is observed for an increasing number of SPFs.
Indeed for small quench amplitudes, see Fig. \ref{convergence} ($a$), the maximum deviation observed in the 
fidelity evolution between the 9 and 12 orbital cases is of the order of $3\%$ for large evolution times $t>225$ that is after the fifth pulse. 
As expected, the case of a larger quench amplitude, presented in Fig. \ref{convergence} ($b$), shows a more demanding convergence behavior.   
However, also in this case a decreasing relative error between different approximations for increasing $M$ is illustrated.  
For instance, the maximum deviation observed in the fidelity evolution calculated using 9 and 12 SPFs respectively is of the 
order of $8\%$ for large evolution times ($t>200$). 
Summarizing, it is important to comment that in both of the above-mentioned cases even the calculation with 6 SPFs is not able to quantitatively predict the dynamics. 
For better visibility of the relative error between different approximations within a positive half of the MIQ protocol we show in the corresponding insets 
of Fig. \ref{convergence} the dynamics during the positive half of the third pulse. 
Note here that the same analysis has also been performed for the dynamics in the eight well potential (omitted here for brevity) showing a very similar behavior.  
Another indicator of convergence is the population of the lowest occupied natural orbital, which is kept below $0.01\%$ for all of our simulations.

\section*{Acknowledgments}
The authors gratefully acknowledge funding by the Deutsche Forschungsgemeinschaft (DFG) in the framework of the
SFB 925 ''Light induced dynamics and control of correlated quantum
systems''. S. I. M thank G. M. Koutentakis for fruitful discussions.

{}

\end{document}